\documentclass[aps,prl,twocolumn,longbibliography,nofootinbib,showkeys,amssymb,floatfix,superscriptaddress]{revtex4-2}

\usepackage{latexsym}
\usepackage{amsfonts}
\usepackage{amsmath}
\usepackage{mathrsfs}
\usepackage{color}
\usepackage[pdftex]{graphicx}
\usepackage{mathptmx}      
\usepackage{bm}
\usepackage{enumerate}
\usepackage{subfigure}
\usepackage{listings}
\usepackage{grffile}
\usepackage{multirow}
\usepackage{qcircuit}
\usepackage{hyperref}
\usepackage{mathtools}
\usepackage[normalem]{ulem}

\definecolor{amethyst}{rgb}{0.6, 0.1, 0.6}
\definecolor{DarkGreen}{RGB}{0,100,0}

\newcommand\ba{\mathbf{a}}

\newcommand\bc{\mathbf{c}}

\newcommand*\widebar[1]{%
   \hbox{%
     \vbox{%
       \hrule height 0.5pt 
       \kern0.5ex
       \hbox{%
         \kern-0.1em
         \ensuremath{#1}%
         \kern-0.1em
       }%
     }%
   }%
}

\newcommand\Cac{1}
\newcommand\Cad{2}
\newcommand\Cbc{3}
\newcommand\Cbd{4}

\newcommand\QUAD{{\Delta}}
\newcommand\SCHSH{{S}}

\begin{document}
\title{Model-free inequality for data of Einstein-Podolsky-Rosen-Bohm experiments}

\author{Hans De Raedt}
\email{deraedthans@gmail.com}
\thanks{Corresponding author}
\affiliation{J\"ulich Supercomputing Centre, Institute for Advanced Simulation,
Forschungszentrum J\"ulich, 52425 J\"ulich, Germany}
\affiliation{Zernike Institute for Advanced Materials,
University of Groningen, Nijenborgh 4, 9747AG Groningen, Netherlands}
\author{Mikhail I. Katsnelson}
\affiliation{Radboud University, Institute for Molecules and Materials,
Heyendaalseweg 135, 6525AJ Nijmegen, Netherlands}

\author{Manpreet S. Jattana}
\affiliation{J\"ulich Supercomputing Centre, Institute for Advanced Simulation,
Forschungszentrum J\"ulich, 52425 J\"ulich, Germany}
\affiliation{Modular Supercomputing and Quantum Computing, Goethe University Frankfurt, Kettenhofweg 139, 60325 Frankfurt am Main, Germany}
\author{Vrinda Mehta}
\affiliation{J\"ulich Supercomputing Centre, Institute for Advanced Simulation,
Forschungszentrum J\"ulich, 52425 J\"ulich, Germany}
\affiliation{RWTH Aachen University, 52056 Aachen, Germany}
\author{Madita Willsch}
\affiliation{J\"ulich Supercomputing Centre, Institute for Advanced Simulation,
Forschungszentrum J\"ulich, 52425 J\"ulich, Germany}
\author{Dennis Willsch}
\affiliation{J\"ulich Supercomputing Centre, Institute for Advanced Simulation,
Forschungszentrum J\"ulich, 52425 J\"ulich, Germany}

\author{Kristel Michielsen}
\affiliation{J\"ulich Supercomputing Centre, Institute for Advanced Simulation,
Forschungszentrum J\"ulich, 52425 J\"ulich, Germany}
\affiliation{RWTH Aachen University, 52056 Aachen, Germany}
\author{Fengping Jin}
\affiliation{J\"ulich Supercomputing Centre, Institute for Advanced Simulation,
Forschungszentrum J\"ulich, 52425 J\"ulich, Germany}

\date{\today}

\begin{abstract}
We present a new inequality constraining correlations
obtained when performing Einstein-Podolsky-Rosen-Bohm experiments.
The proof does not rely on mathematical models
that are imagined to have produced the data and is therefore ``model-free''.
The new inequality contains the model-free version of the well-known Bell-CHSH inequality as a special case.
A violation of the latter implies that
not all the data pairs in four data sets can be reshuffled to create quadruples.
This conclusion provides a new perspective on the implications of the violation of Bell-type inequalities by experimental data.

\end{abstract}

\keywords{Einstein-Podolsky-Rosen-Bohm experiments, Bell's theorem, Bell-Clauser-Horn-Shimony-Holt inequalities, data analysis}
\maketitle


The Einstein-Podolsky-Rosen thought experiment was introduced to question the completeness
of quantum theory~\cite{EPR35}. 
Bohm proposed a modified version that employs spin-1/2 objects instead of coordinates and momenta of
a two-particle system~\cite{BOHM51}.
This modified version, which we refer to as the Einstein-Podolsky-Rosen-Bohm (EPRB) experiment,
has been the subject of many experiments~\cite{KOCH67,CLAU78,ASPE82b,WEIH98,CHRI13,HENS15,GIUS15,SHAL15},
primarily focusing on ruling out 
the model for the EPRB experiment proposed by Bell~\cite{BELL64,BELL93}.

\begin{figure}[!htp]
\centering
\includegraphics[width=0.90\hsize]{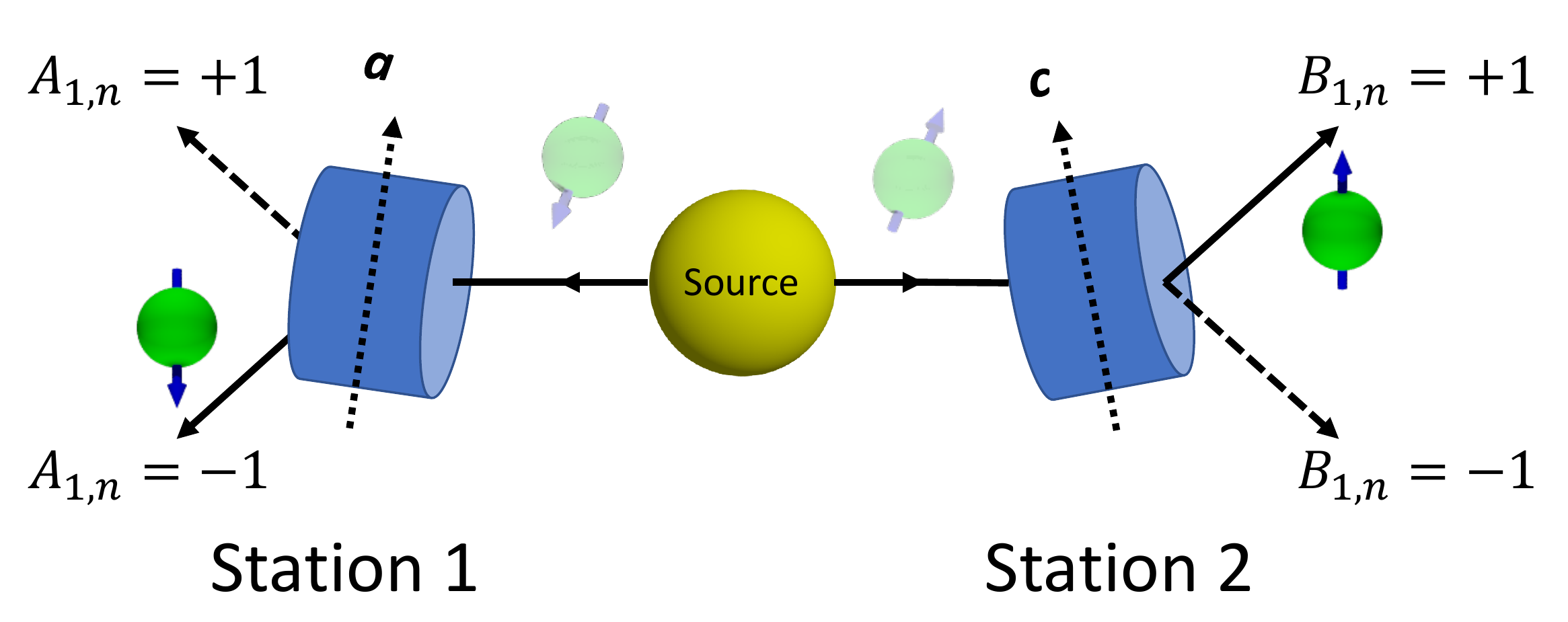}
\caption{(color online)
Conceptual representation of the Einstein-Podolsky-Rosen experiment~\cite{EPR35}
in the modified form proposed by Bohm~\cite{BOHM51}.
A source produces pairs of particles.
The particles of each pair carry opposite magnetic moments, implying that
there is a correlation between the two magnetic moments of each pair leaving the source.
The magnetic field gradients of the Stern-Gerlach magnets (cylinders)
with their uniform magnetic field component
along the directions of the unit vectors $\ba$ and $\bc$ divert each incoming
particle into one of the two spatially separated directions labeled by $+1$ and $-1$.
The pair $(\ba,\bc)$ represents the conditions, denoted by the subscript ``$1$'',
under which the discrete data $(A_{\Cac,n},B_{\Cac,n})$ is collected.
The values of $A_{\Cac,n}$ and $B_{\Cac,n}$ correspond to the labels of the
directions in which the particles have been diverted.
The result of this experiment is the set of data pairs
${\cal D}_{\Cac}=\{ (A_{\Cac,1},B_{\Cac,1}),\ldots, (A_{\Cac,N_1},B_{\Cac,N_1})\}$
where $N_1$ denotes the number of pairs emitted by the source.
The alternative conditions $(\mathbf{a},\mathbf{d})$, $(\mathbf{b},\mathbf{c})$, and $(\mathbf{b},\mathbf{d})$ are labeled by subscripts ``$2$'', ``$3$'', and ``$4$'', respectively.
}
\label{eprbidea}
\end{figure}

The essence of the EPRB thought experiment is shown and described in Fig.~\ref{eprbidea}.
Motivated by the work of Bell~\cite{BELL93} and Clauser et al.~\cite{CLAU69,CLAU74},
many EPRB experiments~\cite{CLAU78,ASPE82b,WEIH98,CHRI13,HENS15,GIUS15,SHAL15}
focus on demonstrating a violation of the Bell-CHSH inequality~\cite{CLAU69,BELL93}.
To this end, one performs four EPRB experiments
under conditions
defined by the directions $(\mathbf{a},\mathbf{c})$,
$(\mathbf{a},\mathbf{d})$, $(\mathbf{b},\mathbf{c})$, and $(\mathbf{b},\mathbf{d})$,
yielding the data sets of pairs of discrete data
\begin{eqnarray}
{\cal D}_{s}&=&\{(A_{s,n},B_{s,n})\,|\,A_{s,n},B_{s,n}=\pm1\,;\,n=1,\ldots,N_s\}
\;,
\label{DATA0}
\end{eqnarray}
where
$s=\Cac,\Cad,\Cbc,\Cbd$ labels
the four alternative conditions and
$N_s$ is the number of pairs emitted by the source.
Then one computes 
correlations according to
\begin{equation}
C_{s}= \frac{1}{N}\sum_{n=1}^N A_{s,n}B_{s,n}
\;,
\label{CORR}
\end{equation}
where $N=\min(N_1,N_2,N_3,N_4)$.


In general, each correlation $C_s$
may take values  $-1$ or $+1$, independent of the values taken by other contributions,
yielding the trivial bound
$|C_{\Cac}\mp C_{\Cad}|+|C_{\Cbc}\pm C_{\Cbd}|\le4$.
Without introducing a specific model for the process generating the data,
we can derive a nontrivial bound that is sharper 
by exploiting the commutative property of addition.


Let $K_{\mathrm{max}}$ be the maximum number of quadruples $(x_k,y_k,z_k,w_k)$ that can be found
by searching for permutations $P$, ${\widehat P}$, ${\widetilde P}$, and $P'$ of $\{1,2,\ldots,N\}$
such that for $k=1,\ldots,K_{\mathrm{max}}$,
\begin{align}
x_k&=A_{\Cac,P(k)}=A_{\Cad,{\widehat P}(k)}
\;,&\;
y_k&=A_{\Cbc,{\widetilde P}(k)}=A_{\Cbd,P'(k)}
\;,
\nonumber \\
z_k&=B_{\Cac,P(k)}=B_{\Cbc,{\widetilde P}(k)}
\;,&\;
w_k&=B_{\Cad,{\widehat P}(k)}=B_{\Cbd,P'(k)}
\;.
\label{QUAD3}
\end{align}
These quadruples are found by rearranging/reshuffling the
data in ${\cal D}_{\Cac}$, ${\cal D}_{\Cad}$, ${\cal D}_{\Cbc}$, and ${\cal D}_{\Cbd}$
without affecting the value of the correlations
$C_{\Cac}$, $C_{\Cad}$, $C_{\Cbc}$, and $C_{\Cbd}$.

{\bf Theorem:} For any (real or computer or thought) EPRB experiment,
the correlations Eq.~(\ref{CORR})
computed from the four data sets
${\cal D}_{1}$, ${\cal D}_{2}$, ${\cal D}_{3}$, and ${\cal D}_{4}$,
must satisfy the model-free inequalities
\begin{eqnarray}
{\cal C}_\pm=\left|C_{\Cac}\mp C_{\Cad}\right|+\left|C_{\Cbc}\pm C_{\Cbd}\right| \le 4-2\QUAD
\;,
\label{DATA4}
\end{eqnarray}
where $0\le\QUAD=K_{\mathrm{max}}/N\le 1$.

{\bf Proof: }We rewrite the correlations in Eq.~(\ref{CORR}) as
\begin{align}
C_{\Cac}&=\frac{1}{N}\sum_{n=1}^N A_{\Cac,{P}(n)}B_{\Cac,{P}(n)},&\;
C_{\Cad}&=\frac{1}{N}\sum_{n=1}^N A_{\Cad,{\widehat P}(n)}B_{\Cad,{\widehat P}(n)}\;,
\nonumber \\
C_{\Cbc}&=\frac{1}{N}\sum_{n=1}^N A_{\Cbc,{\widetilde P}(n)}B_{\Cbc,{\widetilde P}(n)},&\;
C_{\Cbd}&=\frac{1}{N}\sum_{n=1}^N A_{\Cbd,P'(n)}B_{\Cbd,P'(n)}
\;.
\label{QUAD0}
\end{align}
Obviously, reordering the terms of the sums does not change the value of the sums themselves.
As $|x_k|=|y_k|=|z_k|=|w_k|=1$, we (trivially) have $|x_k z_k - x_k w_k| + |y_k z_k + y_k w_k|=2$.

Splitting each of the sums in Eq.~(\ref{QUAD0}) into a sum over $k=1,\ldots,K_{\mathrm{max}}$ and
the sum of the remaining terms, application of the triangle inequality yields
\begin{eqnarray}
{\cal C}_\pm&\le&
\frac{2K_{\mathrm{max}}}{N}
\nonumber \\
&&+
\frac{1}{N}
\sum_{n=K_{\mathrm{max}}+1}^N
\left(
\left\vert A_{\Cac,{P}(n)}B_{\Cac,{P}(n)}\right\vert
+\left\vert A_{\Cad,{\widehat P}(n)}B_{\Cad,{\widehat P}(n)}\right\vert
\right.
\nonumber \\
&&
\hbox to 2cm{}\left.
+\left\vert A_{\Cbc,{\widetilde P}(n)}B_{\Cbc,{\widetilde P}(n)}\right\vert
+ \left\vert A_{\Cbd,P'(n)}B_{\Cbd,P'(n)}\right\vert
\right)
\nonumber \\
&\le&
\frac{2K_{\mathrm{max}}}{N}+\frac{4(N-K_{\mathrm{max}})}{N}= 4 - 2\QUAD
\;.
\label{QUAD4}
\label{DISD7}
\end{eqnarray}
QED.

In the same manner, one can prove a model-free version of the Clauser-Horne inequality~\cite{CLAU74,EBER93}, applicable to the data of the EPRB experiments reported in Ref.~\onlinecite{GIUS15,SHAL15}.

The choice represented by Eq.~(\ref{QUAD3}) is motivated by the EPRB experiment, see Fig.~\ref{eprbidea}.
In general, other choices to define quadruples
are possible and may yield different values of
the maximum fraction of quadruples.

We introduce the Bell-CHSH-like function
\begin{eqnarray}
\SCHSH=\max_{p}
\left|C_{p(1)}-C_{p(2)}+C_{p(3)}+C_{p(4)}\right|
\;,
\label{CHSHdef}
\end{eqnarray}
where the maximum is over all permutations $p$ of
$\{\Cac,\Cad,\Cbc,\Cbd\}$. The maximum guarantees that we cover all possible expressions of the original Bell-CHSH
function~\cite{CLAU69,BELL71,BELL93}.
By application of the triangle inequality, it directly follows from Eq.~(\ref{DATA4}) that, in the case of data
collected by EPRB experiments,
\begin{eqnarray}
\SCHSH\le 4 - 2\Delta
\;.
\label{CHSHineq}
\end{eqnarray}
The symbol $\QUAD$ in Eqs.~(\ref{DATA4}) and~(\ref{CHSHineq}) quantifies structure
in terms of quadruples which can be created by relabeling the pairs of data in the sets
${\cal D}_{\Cac}$, ${\cal D}_{\Cad}$, ${\cal D}_{\Cbc}$, and ${\cal D}_{\Cbd}$.
If $\QUAD=0$, it is impossible to find a reshuffling that yields even one quadruple.
If $\QUAD=1$, the four sets can be reshuffled such that they can be viewed as being
generated from $N$ quadruples. In the special case $\QUAD=1$, we recover the model-free version of the Bell-CHSH inequality
\begin{eqnarray}
\SCHSH\le 2
\;.
\label{CHSH}
\end{eqnarray}
Traditionally, Eq.~(\ref{CHSH}) is proven by assuming that the data can be modeled by
a so-called ``local realistic'' (Bell) model~\cite{CLAU69,BELL71,CLAU74,EBER93,BELL93}.
However, this proof does not extend to the much more general case of the data generated by
EPRB experiments, in contrast to the proof
of Eq.~(\ref{CHSH}) which holds for experimental data.

The proof of the model-free inequalities Eqs.~(\ref{DATA4}) and~(\ref{CHSHineq}) only requires
the existence of a maximum number of quadruples, the actual value of this maximum
being irrelevant for the proof.
However, it is instructive to write a computer program that
uses pseudo-random numbers to generate the data sets
${\cal D}_{\Cac}$, ${\cal D}_{\Cad}$, ${\cal D}_{\Cbc}$, and ${\cal D}_{\Cbd}$
and finds the number of quadruples.
We have implemented the computer program in Mathematica\textsuperscript{\textregistered}.

Naively, finding the value of $\QUAD$ seems to require ${\cal O}(N!^4)$ operations.
Fortunately, the problem of determining the fraction of quadruples $\QUAD$ can
be cast into an integer linear programming problem which is readily solved
by considering the associated linear programming problem with real-valued
unknowns. In practice, we solve the latter by standard optimization
techniques~\cite{PRES03}. For all cases that we have studied, the solution of
the linear programming problem takes integer values only. Then the
solution of the linear programming problem is also the solution of the integer
programming problem.

The results of several numerical experiments using $N=1000 000$ pairs per data set can be summarized as follows:
\begin{itemize}
\item
If the $A$'s and $B$'s are generated in the form of quadruples, all items taking random values $\pm1$,
the program returns $\QUAD=1$,  $|C_{\Cac}-C_{\Cad}|+|C_{\Cbc}+C_{\Cbd}|=0.00063$, and
$|C_{\Cac}+C_{\Cad}|+|C_{\Cbc}-C_{\Cbd}|=0.0028$
such that Eq.~(\ref{DATA4}) is satisfied.
\item
If all $A$'s and $B$'s take independent random values $\pm1$,
the $C$'s are approximately zero. We have $|C_{\Cac}\mp C_{\Cad}|+|C_{\Cbc}\pm C_{\Cbd}|\approx0\le4-2\QUAD$
where in our particular numerical experiment $\QUAD=0.998$.
\item
If the pairs $(A_{\Cac,i},B_{\Cac,i})$,
$(A_{\Cad,j},B_{\Cad,j})$,
$(A_{\Cbc,k}B_{\Cbc,k})$, and
$(A_{\Cbd,l}B_{\Cbd,l})$
are generated randomly with frequencies
$(1-c_{\Cac}A_{\Cac,i}B_{\Cac,i})/4$,
$(1-c_{\Cad}A_{\Cad,j}B_{\Cad,j})/4$,
$(1-c_{\Cbc}A_{\Cbc,k}B_{\Cbc,k})/4$, and
$(1-c_{\Cbd}A_{\Cbd,l}B_{\Cbd,l})/4$, respectively,
the simulation mimics the case of the correlation of two spin-1/2 objects in the singlet state
if we choose $c_{\Cac}=-c_{\Cad}=c_{\Cbc}=c_{\Cbd}=1/\sqrt{2}$.
In this particular case, quantum theory yields
$\max\left(|C_{\Cac}- C_{\Cad}|+|C_{\Cbc}+ C_{\Cbd}|,|C_{\Cac}+ C_{\Cad}|+|C_{\Cbc}- C_{\Cbd}|\right)=2\sqrt{2}\approx2.83$~\cite{CIRE80}.

Generating four times one million independent pairs, we obtain
$\QUAD\approx0.585$,
$\SCHSH=|C_{\Cac}-C_{\Cad}|+|C_{\Cbc}+C_{\Cbd}|\approx2.83$ and $4-2\QUAD\approx2.83$,
demonstrating that the value of the quantum-theoretical upper bound $2\sqrt{2}$
is reflected in the maximum fraction of quadruples that one can create by reshuffling the data.

\item
In the case of Bell's model, slightly modified to comply with Malus' law, 
we have $C_\Cac=-\ba\cdot\bc/2$ and $\SCHSH\le\sqrt{2}$.
Choosing $c_{\Cac}=-c_{\Cad}=c_{\Cbc}=c_{\Cbd}=1/(2\sqrt{2})$
and generating four times one million independent pairs, we obtain
$S=|C_{\Cac}-C_{\Cad}|+|C_{\Cbc}+C_{\Cbd}|\approx1.42$ and $4-2\QUAD\approx2.00$,
as expected for Bell's local realistic model.
\end{itemize}
Except for the first case, the numerical values of $\QUAD$ quoted
fluctuate a little if we repeat the $N=1000 000$ simulations with different random numbers.
In the third case, the simulations suggest that inequality Eq.~(\ref{DATA4}) can be saturated.

Suppose that the (post-processed) data of an EPRB laboratory experiment
yield $\SCHSH>2$, that is the data violate inequality Eq.~(\ref{CHSH}).
From Eq.~(\ref{CHSHineq}), it follows that $\QUAD\le 2-\SCHSH/2<1$ if $\SCHSH>2$.
Therefore, if $\SCHSH>2$ not all
the data in ${\cal D}_{\Cac}$, ${\cal D}_{\Cad}$, ${\cal D}_{\Cbc}$,
and ${\cal D}_{\Cbd}$ can be reshuffled such that they form quadruples only.
Indeed, the data produced by these experiments have to comply with Eq.~(\ref{CHSHineq}) {\color{black}which follows from Eq.~(\ref{DATA4}) holding for data},
and certainly do not have to comply with the original Bell-CHSH inequality obtained from Bell's model.

In other words, all EPRB experiments which have been performed and may be performed in the future
and which only focus on demonstrating a violation of Eq.~(\ref{CHSH}) merely provide evidence
that not all contributions to the correlations can be reshuffled to form quadruples (yielding $\QUAD<1$).
These violations do not provide any clue about the nature of the physical processes that produce the data.

More specifically, Eq.~(\ref{DATA4}) holds for discrete data, irrespective of how
the data sets ${\cal D}_{\Cac}$, ${\cal D}_{\Cad}$, ${\cal D}_{\Cbc}$, and ${\cal D}_{\Cbd}$ were obtained. Inequality~(\ref{DATA4}) shows that correlations of discrete data violate the
Bell-CHSH inequality Eq.~(\ref{CHSH}) only if not all
the pairs of data in Eq.~(\ref{CORR}) can be reshuffled to create quadruples.
The proofs of Eq.~(\ref{DATA4}) and Eq.~(\ref{CHSH}) do not refer to notions
such as ``locality'', ``realism'', ``non-invasive measurements'', ``action at a distance'',
``free will'', ``superdeterminism'', ``(non)contextually'', ``complementarity'', etc.
Logically speaking, a violation of Eq.~(\ref{CHSH}) by experimental data cannot be used to argue about the relevance of one or more of these notions to the process that generated the experimental data.

The existence of the divide between the realm of experimental EPRB data and mathematical models thereof
is further supported by Fine's theorem~\cite{FINE82a,FINE82b}. 
Of particular relevance to the present discussion is the part of the theorem
that establishes the Bell-CHSH inequalities (plus compatibility) as being the necessary and sufficient conditions
for the existence of a joint distribution of the four observables involved in these inequalities.
This four-variable joint distribution returns
the pair distributions describing the four EPRB experiments required to test for a violation of these inequalities.
Fine's theorem holds in the realm of mathematical models only.
Only in the unattainable limit of an infinite number of measurements (that is by leaving the realm of experimental data),
and in the special case that the Bell-CHSH inequalities hold,
it may be possible to prove the equivalence between the model-free inequality
Eq.~(\ref{CHSH}) and the Bell-CHSH
inequality~\cite{CLAU69,BELL71,CLAU74,BELL93}.

A violation of the {\color{black}original (non model-free) Bell-CHSH inequality $S\le2$}
may lead to a variety of
conclusions about certain properties of the mathematical model for which this inequality has been derived.
However, projecting these logically correct conclusions about the mathematical model,
obtained within the context of that mathematical model,
to the domain of EPRB laboratory experiments requires some care, as we now discuss.

The first step in this projection is to feed real-world, discrete data
into {\color{black}the original Bell-CHSH inequality $S\le2$} derived, not for discrete data as we did by considering the case $\QUAD=1$
in Eq.~(\ref{CHSHineq}), but rather in the context of some mathematical model, and to conclude that this inequality is violated.
Considering the discrete data for the correlations as given, it may indeed be tempting
to plug these rational numbers into {\color{black}an expression obtained from some mathematical model}. However, then it is no longer clear
what a violation actually means in terms of the mathematical model because the latter (possibly
by the help of pseudo-random number generators) may not be able to produce these {\color{black} experimental} data at all.
The second step is to conclude from this violation that the mathematical model
cannot produce the numerical values
of the correlations, {\color{black}implying that the mathematical model simply does not apply and has to be replaced by a more adequate one or
that one or more premises underlying the mathematical model must be wrong.}
In the latter case, the final step is to project at least one of these wrong premises to properties of the world around us.

The key question is then to what extent the premises or properties of a mathematical model
can be transferred to those of the world around us.
{\color{black}Based on the rigorous analysis presented in this paper, }the authors' point of view is that in the case of laboratory EPRB experiments, they cannot.


\noindent
\medskip
\begin{acknowledgments}
We are grateful to Bart De Raedt for suggesting that finding the maximum number of quadruples might
be cast into an integer programming problem and for making pertinent comments. We thank Koen De Raedt for many discussions and continuous support.

The work of M.I.K. was supported by the European Research Council (ERC) under the European Union’s
Horizon 2020  research and innovation programme, grant agreement 854843 FASTCORR.  M.S.J. acknowledges support from the project OpenSuperQ (820363) of the EU Quantum Flagship.
V.M., D.W. and M.W. acknowledge support from the project J\"ulich UNified Infrastructure for Quantum computing (JUNIQ) that has received funding from the German Federal Ministry of Education and Research (BMBF) and the Ministry of Culture and Science of the State of North Rhine-Westphalia.


\end{acknowledgments}
\bibliography{all22,bellextra}
\end{document}